\documentstyle[11pt,aaspp4]{article}

\tighten

\lefthead{Rusin et al.}
\righthead{CFHT Imaging of CLASS B1359+154}

\begin{document}

\title{CFHT AO Imaging of the CLASS Gravitational Lens System B1359+154}

\author{D. Rusin\altaffilmark{1}, P.B. Hall\altaffilmark{2,3},
R.C. Nichol\altaffilmark{4}, D.R. Marlow\altaffilmark{1},
A.M.S. Richards\altaffilmark{5}} 

\and

\author{S.T. Myers\altaffilmark{1,6}}

\altaffiltext{1}{Department of Physics and Astronomy, University of
Pennsylvania, 209 S. 33rd St., Philadelphia, PA  19104-6396, USA}
\altaffiltext{2}{Department of Astronomy, University of Toronto, 60 St. George
St., Toronto, ON M5S~3H8, Canada}
\altaffiltext{3}{Visiting Astronomer, Canada-France-Hawaii Telescope,
operated by the National Research Council of Canada, the Centre National
de la Recherche Scientifique de France and the University of Hawaii}
\altaffiltext{4}{Department of Physics, Carnegie Mellon University, 5000
Forbes Ave., Pittsburgh, PA 15232}
\altaffiltext{5}{NRAL, Jodrell Bank, University of Manchester, Macclesfield,
Cheshire SK11 9DL, UK}
\altaffiltext{6}{current address: National Radio Astronomy
Observatory, P.O. Box 0, Socorro, NM, 87801, USA}

\begin{abstract}

We present adaptive optics imaging of the CLASS gravitational lens system 
B1359+154 obtained with the Canada-France-Hawaii Telescope (CFHT) in the
infrared K band. The observations show at least three brightness peaks
within the ring of lensed images, which we identify as emission from multiple
lensing galaxies. The results confirm the suspected compound nature of the
lens, as deduced from preliminary mass modeling. The detection of several
additional nearby galaxies suggests that B1359+154 is lensed by the compact
core of a small galaxy group. We attempted to produce an updated lens model
based on the CFHT observations and new 5-GHz radio data obtained with the
MERLIN array, but there are too few constraints to construct a realistic model
at this time. The uncertainties inherent with modeling compound lenses make
B1359+154 a challenging target for Hubble constant determination through the
measurement of differential time delays. However, time delays will
offer additional constraints to help pin down the mass model.
This lens system therefore presents a unique opportunity to directly measure
the mass distribution of a galaxy group at intermediate redshift.  

\end{abstract}


\section{Introduction} \label{intro}

The gravitational lens system CLASS B1359+154 (Myers et al.\ 1999) was
discovered in the third  phase of the Cosmic Lens All-Sky Survey (Myers et
al.\ 2000), an international program that has imaged over 15,000 flat-spectrum
radio sources at 8.5~GHz using the Very Large Array (VLA) in A-configuration.
The survey seeks to find new cases of gravitational lensing for a variety of
cosmological studies. B1359+154 was identified as a lens directly from
its VLA snapshot map, which featured four compact radio components arranged in
a configuration indicative of lensing. Subsequent optical observations of the
system with the Keck II telescope yielded a redshift of $z=3.235$ for the
lensed background quasar. No spectral features due to the lensing
object were identified. Preliminary mass modeling based on the positions and
flux density ratios of the four lensed radio components strongly suggested
that the lens would be compound, as single-deflector models provided extremely
poor fits to the observables. The system was imaged with the
Canada-France-Hawaii Telescope (CFHT) to investigate this claim. 

In this letter we first summarize the radio data, including a new high
resolution 5-GHz map obtained with the Multi-Element Radio-Linked
Interferometer Network (MERLIN). We then present a CFHT infrared image of
B1359+154, taken using adaptive optics, which conclusively demonstrates that 
the lensing mass is distributed among multiple galaxies. Next, new mass
models based on the above observations are explored. We conclude by discussing
the implications of the CFHT data, along with suggestions for future  studies
of this very interesting lens system. 

\section{Radio Properties} \label{radio}

CLASS B1359+154 was discovered during the CLASS--3 observations of March
1998. Its VLA survey snapshot map showed four compact components (A--D)
arranged in a configuration indicative of a quadruply-imaged source, plus a
fifth extended component inside the four-image configuration. 
A deep VLA follow-up observation at 8.5~GHz resolved the fifth component into
a pair of subcomponents (E and F). Subsequent VLA observations at 15~GHz were
performed to study the spectral properties of the system. The spectral index
of component E between 8.5 and 15~GHz ($\alpha_{8.5}^{15} \approx
-0.6 \pm 0.2$, where $S \propto \nu^{\alpha}$) was found to be flatter than
that of the outer four components ($\alpha_{8.5}^{15} \approx -0.9 \pm
0.1$). This suggested that E was not an additional lensed image, but rather
emission from a weak AGN associated with one of the lensing
galaxies. Component F was not detected at 15~GHz. The upper limit placed on
its spectral index ($\alpha_{8.5}^{15} < -0.6$) could not rule out the
possibility of it being a fifth lensed image. 

MERLIN 5-GHz observations of B1359+154 were performed 1999 January 26 and
February 6. The two observations gave a total integration time of 24
hours on-source. The combined data set was calibrated in AIPS and subsequently
imaged and modelfitted with Gaussian components in DIFMAP (Shepherd 1997). The
final map is displayed in Fig.\ 1 and has a 60 mas beam size. The RMS noise
level in the map is 40 $\mu$Jy/beam. The map shows all six radio components
previously detected with the VLA. Each component was modeled by a single
compact Gaussian. There were hints of a very weak flux bridge joining E and F
at the 2$\sigma$ level in the residual map. Table 1 lists the
component positions, flux densities and spectral indices between $5$ and $8.5$
GHz. We estimate the uncertainties in the spectral indices to be $0.2$ for
A--E and $0.5$ for F. Therefore we cannot dismiss the possibility that
component F is an additional lensed image based on its radio
spectrum. Interestingly, component D appears to have a flatter spectrum
between $5$ and $8.5$ GHz than A--C. This discrepancy may be due to component
variability. To test this assumption, we reduced each of the two MERLIN data
sets separately. We found tentative evidence for variability in the lensed
components at the 10\% level. 

It should be noted that the distribution of flux densities among the B1359+154
lensed radio components is quite peculiar. Based on studies of isolated
lens galaxies, we would expect one of the merging images (B and C) to be
the most highly magnified. However, component A is consistently brighter than
B or C in our radio maps. 

\section{CFHT Observations} \label{cfht}


Infrared observations of B1359+154 using the CFHT were performed 1999 June
23--25. Because of its close proximity to a bright star, the system was
investigated using the Adaptive Optics Bonnette (known as {\it Pue'o}; Rigaut
et al.\ 1998) with the KIR camera, a 1024$\times$1024 Hg:Cd:Te detector with
platescale 0\farcs0348/pixel. All observations were made
with the $K'$ filter (centered at $2.120$ $\mu$m with a FWHM of $0.340$
$\mu$m) using full AO correction. Exposures were 5 minutes long, taken in a
repeated 9-position dither pattern. Images were flat-fielded using the
difference between dome flats taken with lamps on and off. Each night's
flat-fielded data set was medianed to form a blank sky image which was
then subtracted. Integer-pixel image shifts were determined using the
guidestar and one other star visible on most images. An object mask image
was produced from an initial coadd and a second-pass sky image was
constructed with objects masked out. The final image consists of 3.25 hours of 
exposure coadded without pixel rejection. 

A $4'' \times 4''$ section of the CFHT image, centered on the B1359+154 lens
system, is displayed in Fig. 2. Infrared counterparts to all four lensed
images are clearly detected. In addition, a region of extended emission is
observed inside the ring of images, containing three distinct brightness peaks
(K1--K3). The complexity of this extended emission is suggestive of multiple,
superimposed lens galaxies. Overlaying the CFHT image with the MERLIN map, we
see that component E is coincident with a weaker emission feature located
between K1, K2 and K3. Component F is not associated with any of the observed
brightness centers. A wide-field view ($18'' \times 18''$) of the system is
displayed in Fig. 3. Several additional galaxies (K4 -- K6)
are observed within 10\arcsec\ of B1359+154. Traces of an infrared arc
connecting components A, B and C are also visible in the image.

The CFHT data for B1359+154 are described in Table 1. Because components A,
B, C and K1--K6 have well-defined peaks, their positions may be
approximated by eye from the pixel counts. We estimate an error of two
pixels, or 70 mas, on these positions. The CFHT and
MERLIN positions of images A, B and C are consistent within the observational
uncertainties.  Radio components D and E are also identified with CFHT
emission features. However, these features are rather weak and irregular, and
we did not attempt to estimate their positions.

NICMOS standards (Persson et al.\ 1998) measured in 20\arcsec\ diameter
apertures were used to find a photometric solution to the $K$ system, with
0\fm037 RMS accuracy. The integrated magnitude of the B1359+154 lens system is
$m_{K}= 17\fm7$. Magnitudes for the individual CFHT emission features are
given in Table 1. For A--C and K1--K3, magnitudes were calculated within a
$0\farcs4$ diameter aperture centered on the position of the CFHT emission
peak. Identical apertures were also centered at the positions of radio
components D and E. Because of the presence of significant amounts of extended
emission, as well as the apparent superposition of many of the
components, the derived magnitudes are particularly sensitive to the chosen
aperture size. This problem does not affect K4--K6, which are well isolated
from other emission features. Their magnitudes were therefore calculated with
larger ($1\farcs6$ diameter) apertures, and sky-subtracted using an annulus
located $1\farcs2$ -- $2\farcs0$ from the objects.


\section{Lens Modeling} \label{model}

Preliminary mass modeling results presented in Myers et al.\ (1999) strongly
suggested that multiple deflectors would be necessary to account for the
B1359+154 radio data. Models consisting of a singular isothermal ellipsoid
(SIE; Kormann et al.\ 1994) or an SIE plus external shear resulted in
extremely poor fits ($\chi^2/$NDF $\gg 1000$ and $\chi^2/$NDF $> 10$,
respectively), even after allowing for a large uncertainty of $20\%$ on the
image flux density ratios. These findings essentially ruled out a
single-galaxy lens.  An adequate fit was achieved by introducing a second
deflector, parametrized as a singular isothermal sphere (SIS). The CFHT data
has therefore validated the predication of a compound deflector system.

Constructing an improved mass model from the CFHT observations is difficult 
since it is impossible to be sure how many distinct
galaxies reside within the Einstein ring. For example, one of the brightness
peaks may be a star-forming region or disk disturbed by interactions
rather than a galaxy center. Simply fixing deflectors to each of the emission 
peaks is therefore premature. On the other hand, because there are merely 11
constraints (8 positional coordinates and 3 flux density ratios) supplied by
the MERLIN data, only relatively simple compound lens models can be explored if
the positions of the deflectors are left as free parameters. 

Though more and better constraints are needed to properly model this system,
we tested a variety of plausible modeling scenarios based on the
locations of the CFHT brightness peaks and the MERLIN radio data. We allowed
for generous uncertainties of $5$ mas on the positions and $20\%$ on the flux
density ratios of the lensed radio components, and assumed a lens redshift of
$z_l = 1$. 

We first attempted to update the SIE+SIS model by forcing each deflector to
reside within the Einstein ring, without actually fixing their centers to any
observed emission peak. While this model fits the radio data well, it predicts
a fifth lensed image that is significantly above the $3\sigma$ detection limit
of the MERLIN map. The model must be discarded as this image is not observed.
More complex models can be explored by fixing some subset of the deflectors to
the infrared brightness centers (K1--K3, E). We investigated a series of
two-SIE models in which one of the deflectors was fixed while the other was
allowed to float. Each of the four attempts produced poor fits ($\chi^2$/NDF
$\gg 100$). To study the possibility that more than two galaxies are present,
we tested models consisting of two SIEs and one SIS, each fixed to a brightness
peak. This is the most complicated model that can be constrained given the
radio data. Distributing the deflectors among the four brightness peaks leads
to 12 permutations of this model, all of which failed to produce a sufficient
fit using reasonable parameters.

These attempts suggest that B1359+154 will be one of the most challenging
gravitational lens systems to model. Furthermore, if the galaxies are embedded
in a group or cluster, the lensing potential may be significantly more
complicated than our simple models can account for, due to the presence of
large amounts of dark matter not confined to individual galaxies. This dark
matter would contribute significantly to the lensing properties of the system
through magnification and external shear. Clearly the acquisition of
additional constraints is essential to unraveling the mass distribution in the
B1359+154 deflector. Most importantly, we must determine how many galaxies lie
within the lensing potential and pin down their precise locations. Additional
constraints may also be obtained from the images, such as correlated
milliarcsecond substructure in the lensed radio components and the ratios of
measured time delays.

\section{Conclusions and Future Work} \label{conclude}

CFHT AO observations of the CLASS gravitational lens system B1359+154 have
detected infrared counterparts to each of the lensed radio components, as well
as a region of extended emission within the four-image configuration. This
region includes at least three distinct brightness centers, indicative of
multiple, superimposed lensing galaxies. Several additional galaxies are
observed to be flanking the system. As only $1.1 \pm 0.3$ galaxies would be
expected down to $K=21$ per $18'' \times 18''$ field (Hall, Green and Cohen
1998), this represents a significant overdensity. These observations suggest
that the lensing mass in the B1359+154 system corresponds to the compact core
of a small galaxy group. 

The morphology of the K1--K3 emission region, along with the extremely poor
fits provided by single-galaxy lens models, provide compelling evidence for a
compound lens. The detection of coincident radio emission is consistent with
this hypothesis, as AGNs are often triggered by interactions in dense
environments. Interestingly, if there are more than two galaxies within the
Einstein ring, they would each need to be significantly undermassive to
produce an image splitting of only $1\farcs7$.  The lens candidate HST
180746+45599 (Ratnatunga, Griffiths and Ostrander 1999) also appears to be
lensed by a small group of undermassive galaxies, possibly signaling the
importance of such deflectors in arcsecond-scale lensing.

The confirmation of another compound lens system in CLASS seems to
contradict the claim that lenses comprised of more than one primary 
galaxy should contribute negligibly to the overall lensing rate (Keeton,
Kochanek and Seljak 1997). Three of the eleven new gravitational lens systems
discovered in CLASS -- B1359+154, B1127+385 (Koopmans et al.\ 1999) and
B1608+656 (Koopmans and Fassnacht 1999) -- contain multiple galaxies within
the Einstein ring. The lensing potentials of two additional systems --
B1600+434 (Koopmans, De Bruyn and Jackson 1998) and B2319+051 (Marlow et al.\
2000) -- are strongly influenced by large secondary deflectors within a few
arcseconds. Such observations suggest that compound deflectors play a
significant role in lensing, a role that should be investigated with regard to
lensing statistics analyses and constraints on the cosmological parameters.

Deep optical and/or near-infrared observations are needed to properly study the
structure of the B1359+154 deflector system, pin down the number and positions
of all galaxies present, and provide the necessary constraints on the mass
model.  High sensitivity, multi-color imaging with HST or Gemini is essential
to correctly identifying all of the emission features, distinguishing
among galaxy centers, star-forming regions and any disturbed disks or tidal
tails which may have been induced by interactions. Because spectroscopy on
each of the individual components is quite challenging, these observations
will also provide vital photometric redshifts for the system. We should note
that for a source at $z=3.235$, the lens redshift is mostly likely to be 
$z \approx 0.7$ and may be as high as $z\approx 1$. The magnitude of the
isolated galaxy K4 is consistent with an L* elliptical in this redshift range
for a flat $\Lambda$-dominated cosmology. 

Much is still to be learned from high resolution radio investigations of this
system. The detection of correlated milliarcsecond substructure in the
lensed images would provide essential constraints to the mass model. As many as
three additional constraints can be obtained from the ratios of measured
time delays. The nature of radio component F also remains to be determined. At
this time we cannot rigorously rule out the possibility of it being a fifth
lensed image. If there are mass distributions centered at K1 and K3, the
Fermat potential (Schneider, Ehlers and Falco 1992) should have a saddle point
between them in which an image could form. Component F is located in this
general region. Evidence for a faint flux bridge connecting E and F argues
against the fifth-image hypothesis. A deep 1.7-GHz observation with the VLBA
offers the best chance of conclusively detecting this flux bridge, if it
exists. 

B1359+154 has one of the most complex deflectors of any small separation (1--2
arcsec) gravitational lens system, and will be one of the most challenging
lenses to model. Many more observational constraints are needed to
meet this challenge. The construction of a viable lens model will offer a
unique opportunity to probe the mass distribution of a small galaxy group at
intermediate redshift. With the uncertainties inherent in the modeling of
compound lenses, however, B1359+154 is unlikely to be a promising target for
Hubble constant determination.

\acknowledgments

We thank the staff of the CFHT and MERLIN for their assistance during our
observing runs. MERLIN is a national UK facility operated by the University of
Manchester on behalf of PPARC. STM was supported by an Alfred R. Sloan
Fellowship at the University of Pennsylvania. DR thanks Leon Koopmans for his
advice and suggestions during the preparation of this draft.

\def\phb{\phantom{00}}

\clearpage

\begin{table*}
\caption{CLASS B1359+154: MERLIN Positions and Flux Densities; CFHT Positions
and Magnitudes.}
\label{tab1} 
\medskip
{\small
\begin{tabular}{ c c c c c | c c c}
\hline \hline
 & & MERLIN Data & & & & CFHT Data & \\
\hline
Comp & \sc Offset (E) & \sc Offset (N) & $S_5$ (mJy) &
$\alpha_{5}^{8.5}$ & \sc Offset (E) & \sc Offset (N) & $m_{K}$\\ 
\hline
 & & & & & & & \\
A  & $+0\farcs0000 \pm 0\farcs0001$ & $+0\farcs0000 \pm 0\farcs0001$ & $19.0$ &
$-1.3$ & $+0\farcs00$ & $+0\farcs00$ & $20\fm8$ \\
B  & $-0\farcs4915 \pm 0\farcs0002$ & $-1\farcs2480 \pm 0\farcs0002$ & $11.2$ &
$-1.2$ & $-0\farcs52$ & $-1\farcs25$  & $20\fm8$\\
C  & $-0\farcs3137 \pm 0\farcs0002$ & $-1\farcs6640 \pm 0\farcs0002$ & $15.8$ &
$-1.3$ & $-0\farcs38$ & $-1\farcs64$  & $20\fm7$\\
D  & $+0\farcs9571 \pm 0\farcs0008$ & $-1\farcs3660 \pm 0\farcs0008$ & $ 3.2$ &
$-1.0$ & -- & -- & $21\fm2$ \\
E  & $+0\farcs6072 \pm 0\farcs0005$ & $-1\farcs1420 \pm 0\farcs0005$ & $ 4.7$ &
$-1.5$ & -- & -- & $20\fm7$ \\
F  & $+0\farcs4120 \pm 0\farcs0020$ & $-0\farcs9580 \pm 0\farcs0020$ & $ 1.2$ &
$-1.7$ & -- & -- & -- \\
K1 & -- & -- & -- & -- & $+0\farcs63$ &  $-0\farcs70$ & $20\fm7$ \\
K2 & -- & -- & -- & -- & $+0\farcs52$ &  $-1\farcs50$ & $20\fm7$ \\
K3 & -- & -- & -- & -- & $+0\farcs14$ &  $-0\farcs94$ & $20\fm7$ \\
K4 & -- & -- & -- & -- & $+3\farcs62$ &  $-1\farcs36$ & $19\fm1$ \\
K5 & -- & -- & -- & -- & $-1\farcs81$ &  $+1\farcs50$ & $21\fm0$ \\
K6 & -- & -- & -- & -- & $-6\farcs51$ &  $-6\farcs99$ & $21\fm0$ \\

\hline
\end{tabular}}
\tablecomments{MERLIN positions are offset from RA 14 01 35.5495 and Dec +15
13 25.643 (J2000). The rms noise level is $40 \mu$Jy/beam. Spectral indices
between 5 GHz and 8.5 GHz are computed using the deep X-band observation
presented in Myers et al.\ 1999. Uncertainties in the MERLIN positions are
estimated from the beam size (60 mas) divided by the signal-to-noise.  We
estimate the uncertainties in the CFHT positions to be approximately 70 mas.}
\end{table*}

\clearpage
\onecolumn

\begin{figure}
\epsfxsize=5in
\epsfbox[33 -45 582 640]{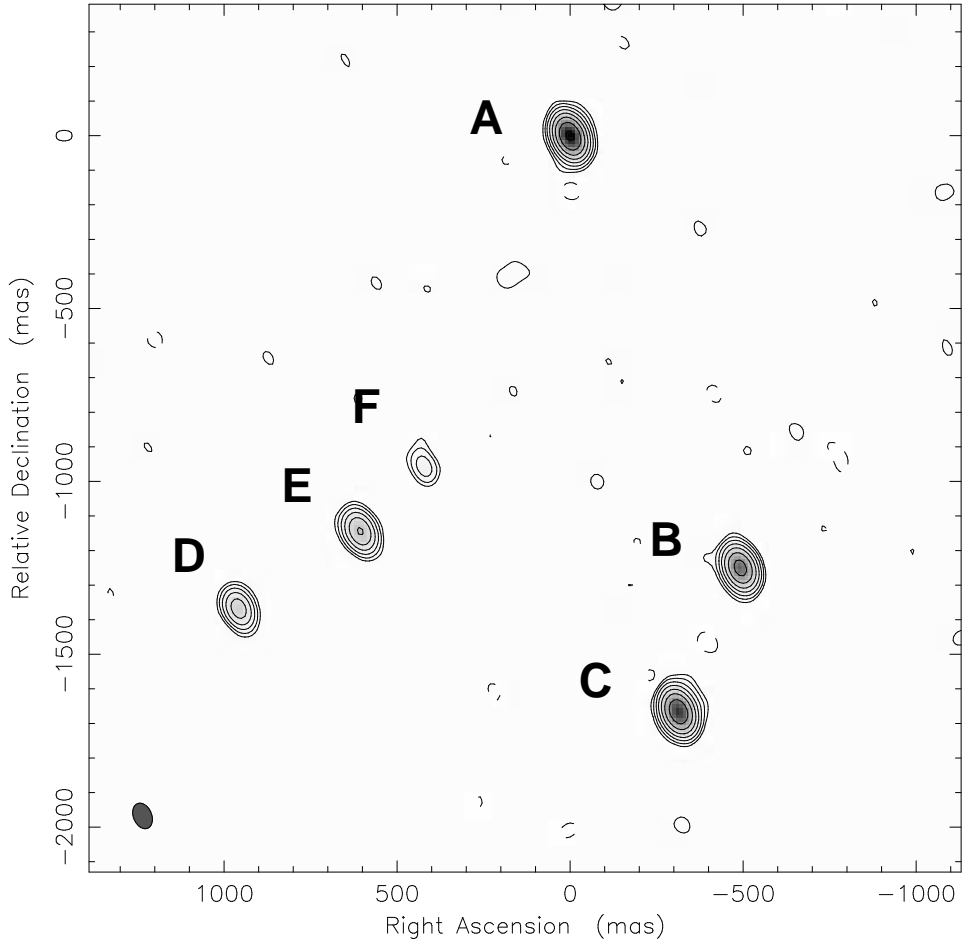}
\figurenum{1}
\caption{MERLIN 5-GHz map of CLASS B1359+154.}
\end{figure}

\begin{figure}
\epsfxsize=5in
\epsfbox[33 -45 582 640]{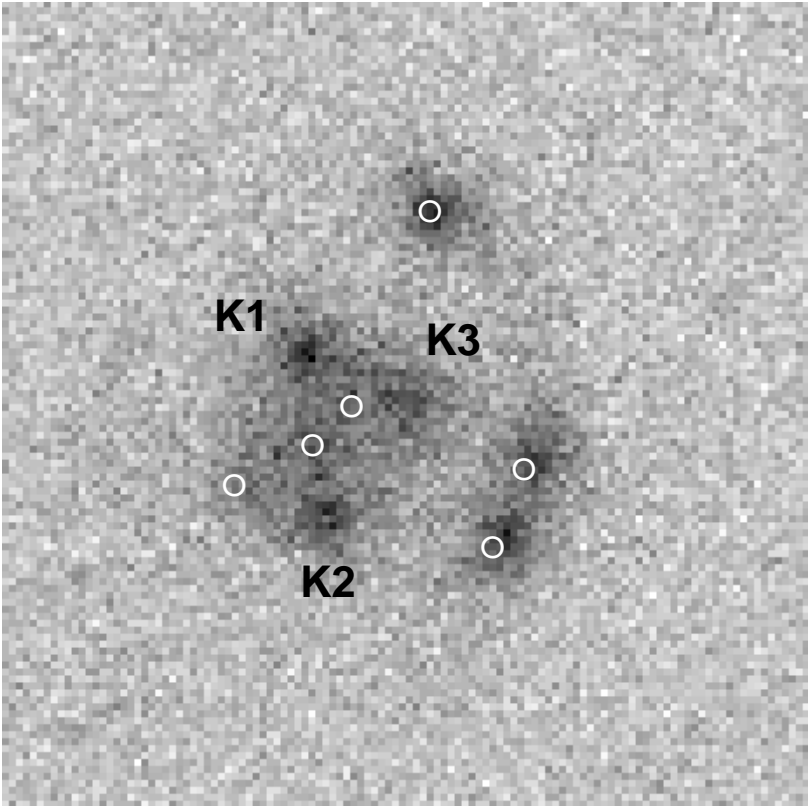}
\figurenum{2}
\caption{CFHT K-band image of CLASS B1359+154 ($4'' \times 4''$). Positions of
the MERLIN radio components are denoted by circles. Radio component A is fixed
to the peak of the corresponding CFHT emission feature.}
\end{figure}

\begin{figure}
\epsfxsize=5in
\epsfbox[33 -45 582 640]{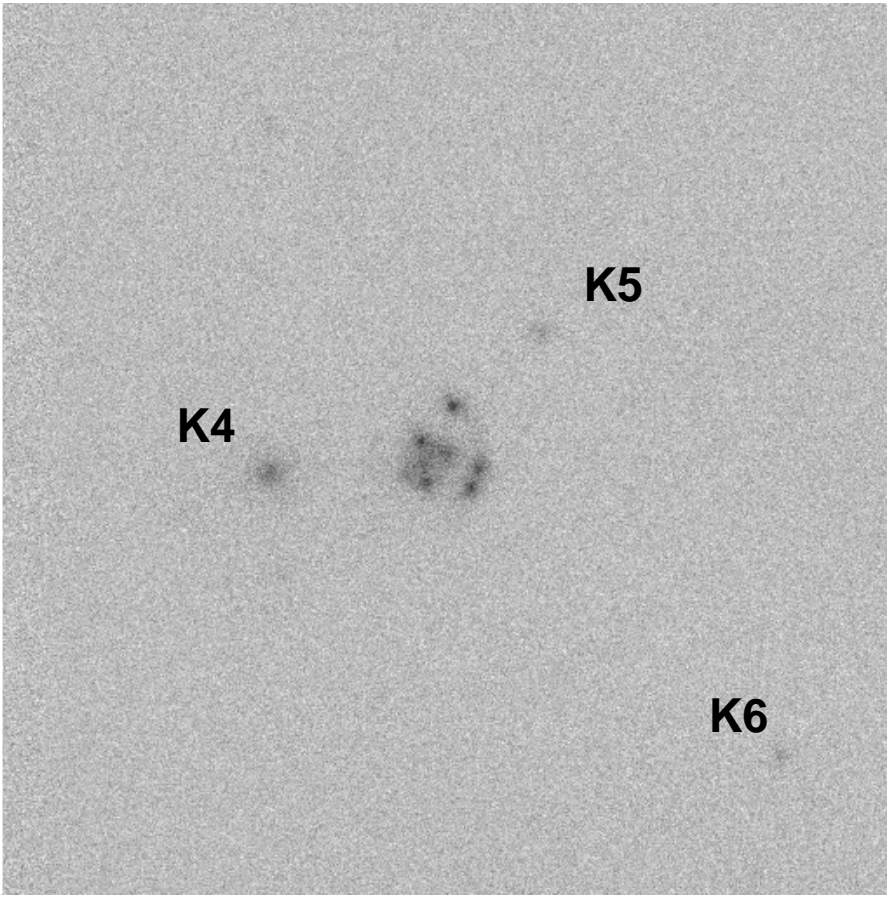}
\figurenum{3}
\caption{CFHT K-band image of CLASS B1359+154 ($18'' \times 18''$).}
\end{figure}

\end{document}